\documentclass[a4page,10pt]{article}
\usepackage[top=0.8in, bottom=0.8in, left=0.5in, right=0.5in]{geometry}
\usepackage{fancyhdr}
\renewcommand{\thispagestyle}[1]{}
\usepackage[utf8]{inputenc}
\usepackage{hyperref}
\usepackage{tabularx}
\usepackage{graphicx}

\usepackage{wrapfig}
\usepackage{xcolor}

\newtheorem{defn}{Definition}[section]

\begin{document}

\title{Efficiently Discovering Frequent Motifs in Large-scale Sensor Data}

\author{Puneet Agarwal, Gautam Shroff, Sarmimala Saikia, and Zaigham Khan\\
{TCS Research, Tata Consultancy Services Ltd., Noida}\\
{\{puneet.a, gautam.shroff, sarmimala.saikia, zaigham.khan\}@tcs.com}}

\maketitle
\begin{abstract}
While analyzing vehicular sensor data, we found that frequently occurring waveforms
could serve as features for further analysis, 
such as rule mining, classification, and anomaly detection. 
The discovery of waveform patterns, also known as time-series \textit{motifs}, has been studied extensively; 
however, available techniques for discovering frequently ocurring time-series motifs were found lacking in either efficiency or quality:
Standard subsequence clustering results in poor quality, to the extent that it has even been termed `meaningless'. 
Variants of hierarchical clustering using techniques for efficient discovery of `exact pair motifs' find high-quality frequent motifs, 
but at the cost of high computational complexity, making such techniques unusable for our voluminous vehicular sensor data.
We show that good quality frequent motifs can be discovered using bounded spherical clustering of time-series subsequences,
which we refer to as COIN clustering, with near linear complexity in time-series size. 
COIN clustering addresses many of the challenges that previously led to subsequence clustering being viewed as
meaningless. We describe an end-to-end motif-discovery
procedure using a sequence of pre and post-processing techniques that remove trivial-matches and shifted-motifs,
which also plagued previous subsequence-clustering approaches. 
We demonstrate that our technique efficiently discovers frequent motifs in voluminous vehicular sensor data  
as well as in publicly available data sets.
\end{abstract}

\section{Introduction}
\label{sec:intro}
Many modern vehicles are fitted with numerous sensors that continuously record a variety of parameters related to their health and usage,
often producing many long time-series for every vehicle. Engineering and quality departments are tasked with analyzing collections
of such time-series data across a large population of vehicles to better understand the behavior of the vehicle model. 

Our broader goal is to discover interesting temporal correlations between events occurring in one or more sensors via temporal rule mining.
In addition to known events, such as `rapid deceleration', we also wanted to include events that might not be a priori known to engineers.
In this context, we sought to discover \textit{frequently} occurring waveform patterns, or \textit{motifs}, within each sensor time-series,
and use these as potential
events for further temporal rule mining. In this paper we focus on our journey of discovering frequent motifs for our collection of
vehicular data from more than two dozen sensors, each being a time-series with more than half a million values. 

We realized that the problem of discovering waveform patterns, or \textit{motifs}, within a single time-series has been extensively studied; 
some of these techniques do focus on finding \textit{frequent} motifs \cite{p:KeoghMotifs, p:epenthesisMotif}, 
while others seek to find the closest pairs of similar subsequences \cite{p:KeoghExactMotifs,p:KeoghImageMotifs}. 
Unfortunately, techniques given in \cite{p:KeoghMotifs} as well as more recently \cite{p:epenthesisMotif} (which relies on iteratively finding the closest pair of
waveforms using \cite{p:KeoghExactMotifs}), are of at least quadratic complexity in time-series size and so were found to be unusable for
the volume of vehicular sensor data we were dealing with. Additionally, focus of our work and that of
\cite{p:epenthesisMotif,p:effiMotif,p:KeoghExactMotifs,p:KeoghImageMotifs} are different as described in later Section \ref{sec:rw}.

Another, possibly more efficient approach to frequent-motif discovery has been to group subsequences of time-series using standard
clustering techniques. However, there have been mixed views about meaningfulness of subsequence time-series (STS) clustering. 
Keogh et al. \cite{p:keoghClustMeaning} demonstrated that STS clustering gives results that are independent of the dataset used; 
thereafter many others \cite{p:ChenICDM,p:ChenKAIS,p:useful,p:tskernelDensityClustering,p:meaningful,p:unfoldingNN} have refuted such claims. 
Further, as also pointed out in \cite{p:tskernelDensityClustering}, these algorithms, which group subsequences using
standard clustering techniques, are of quadratic complexity, and so would also not serve our purpose. 

In practice, we observed that useful patterns do indeed occur frequently in the time-series we were confronted with, and 
they appeared to be representable by clusters of subsequences. So, while the dust has probably settled on
the debate around subsequence clustering, 
we concluded that its potentially better computational efficiency rendered it the most promising approach for our purpose, i.e., analyzing 
large volumes of vehicular data. However, we found that techniques for \textit{efficiently}  discovering motifs in voluminous time-series
(i.e., near linear time in the size of the time-series), using subsequence clustering or otherwise, were neither publicly available nor well addressed
in the literature.

\noindent \textbf{Contribution 1:} In this paper we present significantly more efficient algorithms that group subsequences using bounded spherical
clustering, for which we introduce the term `COIN clustering'. Our algorithms behave near linearly in time-series size, and so
were easily able to process our voluminous vehicular data. 
One of our approaches uses a bounded spherical BIRCH clustering \cite{p:birch}. We also present another COIN clustering approach using
locality-sensitive hashing \cite{b:mmds,p:IndykGaussianLSH} that is potentially more amenable to parallel implementation (however, the parallel
version is not covered here). 

Our approach uses only one parameter, i.e., the window size $w$. This is one more parameter as compared
to the parameter-less approach of \cite{p:epenthesisMotif}. However, we submit that this is a small price to pay for the vastly increased
efficiency: running our algorithms on a range of window sizes will still be cheaper than the technique of \cite{p:epenthesisMotif}.

We also found that clustering subsequences is only a part of the complete frequent-motif discovery problem,
and other important steps were required, such as trivial match removal, and removal of duplicate motifs, i.e., those 
that are merely shifted versions of each other.  

\noindent \textbf{Contribution 2:} We describe an end-to-end motif-discovery
procedure that uses a sequence of pre and post-processing techniques to remove trivial-matches and shifted-motifs.
We also introduce steps that are useful in practical applications, such as `level splitting', which distinguishes between
occurrences of motifs at different levels, while still representing the same waveform shape: For example, it does not makes sense
 to consider a `sudden rise' in temperature from zero to 50 degrees in the same `motif' as a similar, 
equally sudden rise from 100 to 150 degrees; the former may be quite natural, while the latter might result in system failure.  

We ran our end-to-end frequent-motif discovery algorithms on both our voluminous collection of vehicular sensor data
as well as many far smaller publicly available data sets and matched the motifs it discovered with the results of previous work on the same data sets.

The remainder of this paper is organized as follows: We present the background and formal definitions in the next section. In
Section~\ref{sec:sphClust} we motivate the need for spherical clustering to discover frequent motifs. Section~\ref{s:coin} describes
COIN clustering, including those using BIRCH as well as LSH, and Section~\ref{s:motif} explains the end-to-end process for frequent motif
discovery including trivial-match removal and other steps such as `level splitting' that are important in practice.
In Section~\ref{sec:exp} experimental results are analyzed. Finally, we place our contributions in the context of 
related work in Section \ref{sec:rw}, in Section~\ref{sec:applications} we place our work in the larger context 
and briefly indicate how we use the discovered frequent motifs for further analysis, before concluding 
with a discussion in Section~\ref{sec:concl}.
\section{Background}
\label{sec:Def}
The problem of discovering frequent time-series motifs has been extensively discussed in previous research
\cite{p:KeoghMotifs,p:meanShiftMotif,p:KeoghExactMotifs,p:KeoghImageMotifs}. However, our formulation is slightly different and
so we formally define \textit{frequent motifs} in the context of time-series data arising from a single sensor as follows:

Consider a time-series $T_i = \{v_1, v_2, \dots, v_{n_i}\}$ representing values arising from a single sensor sampled at
regular intervals on a temporal scale $t_1 \ldots t_{n_i}$.
In practice there will be many time-series even for a single sensor, $T = \{T_1, \dots, T_m\}$: 
Each $T_i$ arises from a single operation
of one of the underlying systems: in our case, each $T_i$ arises from a continuous run of one vehicle,
and the entire data $T$ consists of data from multiple runs of many vehicles.

We first normalize each series $T_i$ to zero mean and unit variance to get $Z_i = \{z_1, ..., z_j, ..., z_{n_i}\}$, i.e., $z_j$ is the z-score of $v_j$. 
Next we generate subsequences $S_i$ from $Z_i$, using a moving window of length $w$, i.e., 
$S_i$ consists of $n_i - w +1$ subsequences of length $w$, starting at times $t_1 \ldots t_{n_i - w +1}$. 
From $m$ such time-series we get a consolidated set of subsequences $S = \bigcup_{k} S_k$, $k=1 \ldots m$, where $|S| = (\sum_i n_i-mw+m)$. 
We shall refer to $S$ as \textit{subsequence-matrix} for the set of time-series $T = \{T_1, \dots, T_m\}$.

Elements in $S$ are considered to be instances of the same pattern or motif if
they are close to each other in terms of a distance measure $D(\bf{s_1},\bf{s_2})$ on the set of subsequences. 
For most of the discussion in this paper we assume $D$ to be the \textbf{Euclidean distance}
between the two subsequences \textit{after} each sequence is normalized to have zero mean (we shall explain 
the rationale for this normalization in due course).

As also pointed out in earlier work \cite{p:KeoghMotifs,p:keoghClustMeaning}, since $S$ contains a subsequence starting at \textit{every} time step of each particular series $T_i$, 
neighboring subsequences will be trivially close to each other using any reasonable measure including Euclidean distance. 
Non-trivially matching subsequences are defined as in \cite{p:KeoghMotifs,p:keoghClustMeaning}.
\begin{defn}
\label{def:sim}
Two subsequences $\bf{s_1}$ and $\bf{s_2}$ are called non-trivially similar, if there exists a subsequence $\bf{s_d}$ such that
$D(\bf{s_1}, \bf{s_d}) > \delta$, $D(\bf{s_2}, \bf{s_d}) >\delta$, $D(\bf{s_1},\bf{s_2)} \leq \delta$, while $\bf{s_d}$ occurs
 between $\bf{s_1}$ and $\bf{s_2}$, i.e., if $t_1$, $t_2$, and $t_d$ are the start-time of these subsequences, then $t_1 < t_d < t_2$ or
$t_2 < t_d < t_1$. Here, $\delta$ is a prior threshold.
\end{defn}  
\begin{defn}
A frequent motif $M$, $(M \subset S)$ is set of non-trivially similar subsequences of cardinality greater than
a threshold $s$.
\label{def:motif}
\end{defn}
(Note that out of many temporally overlapping subsequences in a frequent motif Definition \ref{def:sim} identifies those that are
most similar to that other non-overlapping instances of the motif.)

\section{Spherical Clustering for Discovering Frequent Motifs}
\label{sec:sphClust}
Clustering time-series subsequences has been studied extensively
\cite{p:ChenICDM,p:ChenKAIS,p:useful,p:KeoghMotifs,p:KeoghExactMotifs,p:epenthesisMotif,p:unfoldingNN,p:tsClustering}.
In particular Keogh et al. \cite{p:keoghClustMeaning} have claimed that subsequence clustering gives meaningless results. One
of the causes for this non-intuitive conclusion has been rightly identified \cite{p:ChenICDM,p:ChenKAIS} to be the fact that time-series
subsequences are highly \textit{correlated}:
Consecutive subsequences in the subsequence-matrix are similar to each other since neighboring values vary slowly in a smooth time-series.
Consequently, subsequence clustering results in `trivial matches' \cite{p:KeoghMotifs,p:keoghClustMeaning} that need to be identified and removed
after clusters are found \cite{p:useful}.

 \begin{wrapfigure}[14]{R}{0.55\columnwidth}
 \includegraphics[width=0.5\columnwidth]{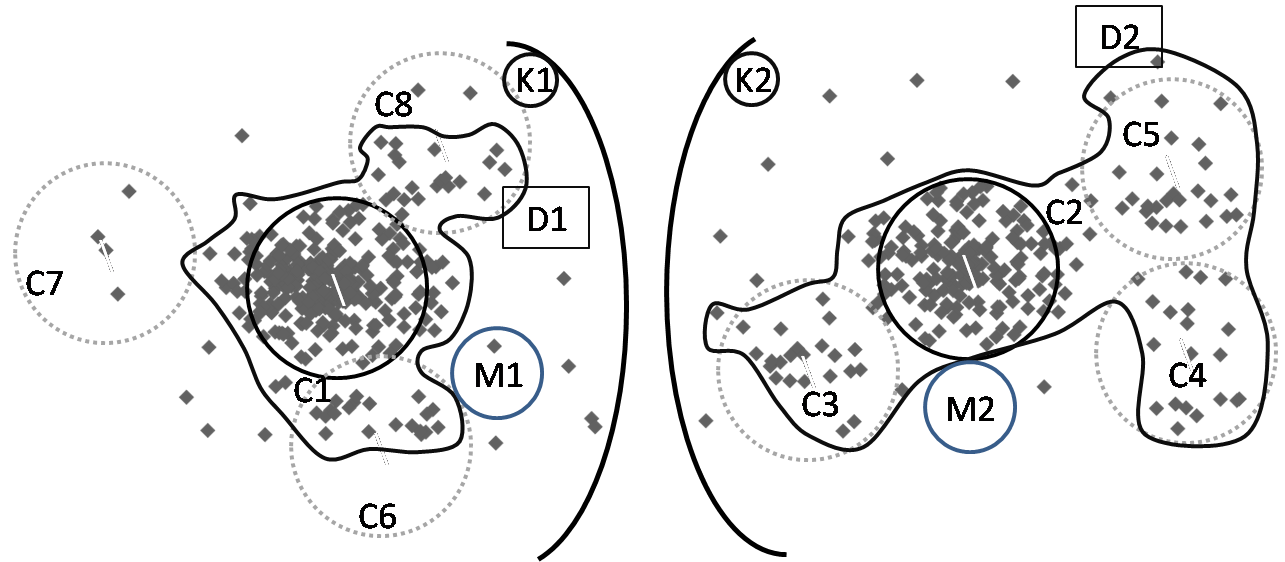}
 \caption{Clustering Techniques}
 \label{fig:clustering}
 \end{wrapfigure}

Further, highly correlated subsequences effectively lie along a long 
lower dimensional manifold in $w$-space. As we traverse the length of this manifold, the successive pairs of points (subsequences) we encounter 
are actually close to each other, which also results in traditional clustering algorithms behaving poorly in segregating
true frequent motifs that form tighter clusters, as we illustrate in Figure \ref{fig:clustering}. For instance, k-means might identify large diameter
clusters $K_1$ and $K_2$ separated by a boundary, and density-based clustering (such as DBSCAN) often results in elongated clusters such as $D_1$ and $D_2$.
In each case pairs of subsequences within a cluster are \textit{not} within a small $\delta$ distance of each other.
So, while the true frequent motifs $M_1$ and $M_2$ may get separated into separate clusters, many extraneous points also naturally accumulate in each cluster due 
to the trivial-match problem, effectively obfuscating the true motifs. Chen also observed the same issue \cite{p:ChenICDM,p:ChenKAIS}.

If on the other hand we use spherical clustering with a hard bound on radius (what we term below as COIN clustering),
we tend to find bounded clusters such as $C_1$ and $C_2$. Of course, the other,
extraneous points are also covered by similar spherical clusters, shown with lighter outlines; however, these are usually
dropped as insignificant based on their lower support (i.e., while there are many such spheres, each has but a few elements).

  \begin{wrapfigure}[19]{R}{0.65\columnwidth}
 \includegraphics[width=0.6\columnwidth]{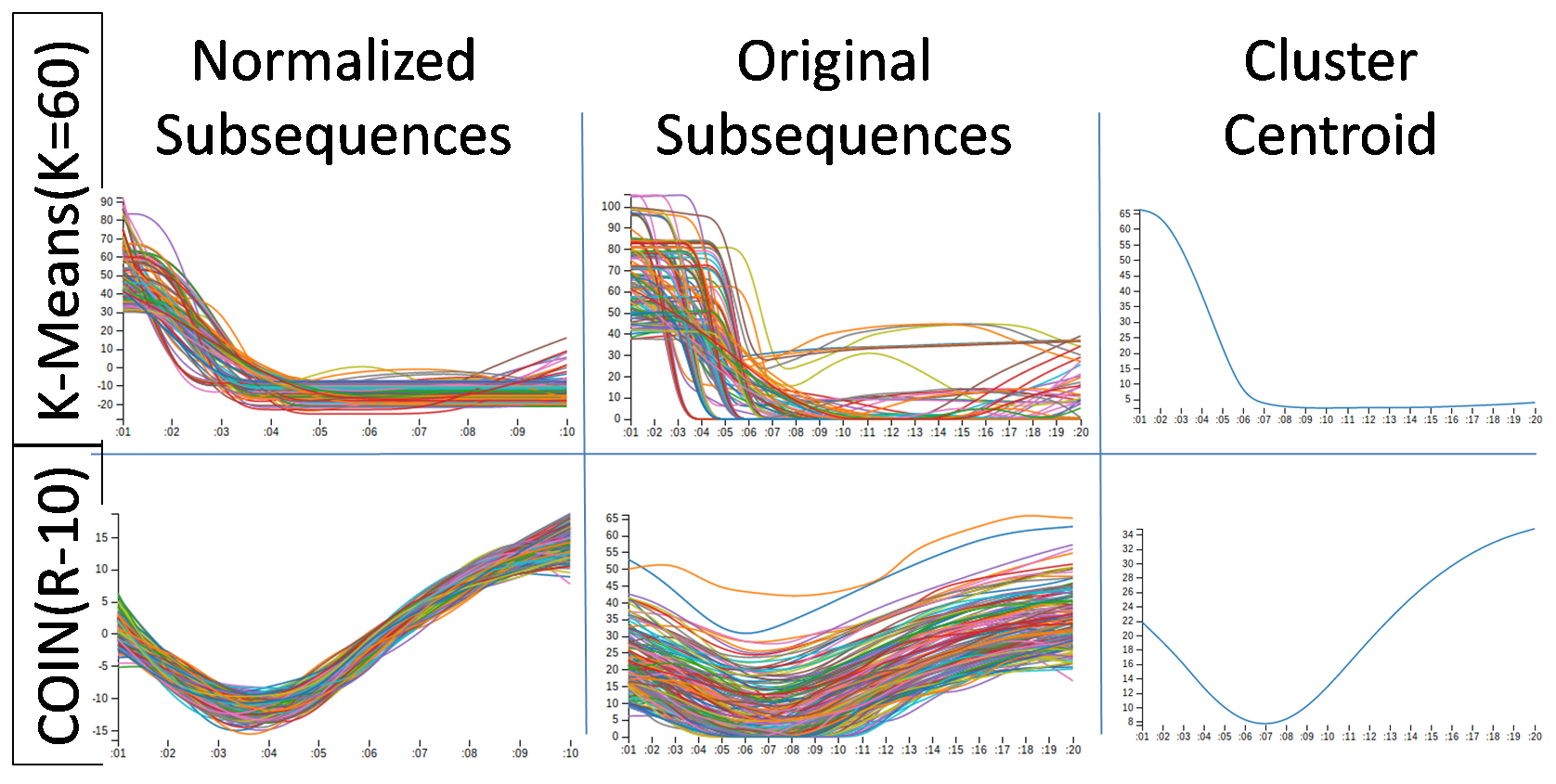}
 \caption{k-means vs spherical COIN clustering, showing time on x-axis, and sensor value on y-axis. Exact values on the axes are not
important.}
 \label{fig:whySph}
\end{wrapfigure}

Figure \ref{fig:whySph} shows the result of clustering subsequences using k-means versus one of our COIN techniques and illustrates the above phenomenon.
While the motif identified by COIN shows a tight pattern, the corresponding k-means cluster includes many extraneous points.
Note that while clustering is performed on the normalized subsequences, the COIN motif's distinctive shape is visible even in the original sequence
 (i.e., before mean-shifting), while
it is completely obscured by noise when k-means is used. Also, the COIN cluster centroid is indeed a representative shape
for the actual pattern while this is not the case using k-means (as also observed by Keogh et al. in
\cite{p:keoghClustMeaning} and Chen in \cite{p:ChenICDM,p:ChenKAIS}).

Another important point is that techniques such as k-means become increasingly inefficient for large $k$. In our practical application with vehicular sensor
data, time-series often contain thousands if not tens of thousands of clusters that need to be discovered, even if only a few hundred are significant in terms
of support. Thus the number of clusters that need to be sought to find tight motifs is large and also varies greatly from sensor to sensor. In such situations
techniques such as k-means appear inappropriate, both due to time complexity as well as the difficulty in choosing the right value of $k$ 
efficiently.

Last but not least, it is important to note that spherical clustering of time-series subsequences can and usually does result in \textit{overlaps},
i.e., many subsequences end up close to more than one cluster center. This is a natural consequence of the nature of the space
of time-series subsequences rather than any reflection on the clustering technique. The important point is that spherical clustering 
manages to find the true frequent motifs, as we shall demonstrate via experimental results on real-life data, as well as on public datasets
where ground truth is known.

We have used a well known COIN clustering technique called BIRCH \cite{p:birch} that has been shown to
outperform standard clustering techniques on large data sets \cite{p:clusteringSurvey1}. 
As we shall show in Section \ref{sec:exp}, using BIRCH is also significantly more efficient than techniques
such as Epenthesis \cite{p:epenthesisMotif} for discovering frequent time-series motifs.
Additionally, we also present another equally efficient COIN clustering technique, based on locality-sensitive hashing that
is better amenable for parallel implementation.

Finally, even after trivial matches have been removed from each potential true motif discovered as a high-support spherical cluster,
it is still possible that a single frequent motif (usually one with rather high support) finds itself split into multiple spherical clusters: Multiple
such spherical clusters capture \textit{shifted} versions of the true motif. So we finally detect and remove such duplicates
by comparing the starting points of subsequences in pairs of high-support clusters. 
\begin{defn}
 \label{def:shiftedClust}
 Two clusters of subsequences $C_1,  C_2$ (where $|C_1| \leq |C_2|$) are deemed to be shifted versions of each other if at-least p\% of the subsequences of $C1$ match trivially with the subsequences of the cluster $C_2$.
\end{defn}

As a result of these multiple steps, viz. subsequence clustering,
duplicate elimination and finally removing trivial matches, we manage to find largely unique frequent motifs. At the same time,
it is important to note that multiple runs of the same process can result in \textit{different} patterns modulo shifts. 
So, in terms of the measures described by Keogh et al. in \cite{p:keoghClustMeaning}, the motifs are non-repeatable, which
was also one of the reasons for their being called meaningless. However, the fact is that we are able to find true
frequent motifs, albeit modulo shifts, does indeed suffice for all practical purposes for vehicular sensor data.

\section{COIN Clustering}
\label{s:coin}
We introduce the term \textit{COIN clustering} to describe techniques that result in spherical clusters 
that bound the maximum distance between any two members of a single cluster to be less than a fixed cluster diameter, $2R$.
COIN clustering uses the pictorial analogy of a `coin' to perform clustering (though `sphere' would be more correct),
with the radius of the coin being the \textit{distance threshold} $R$. 
Our goal is to ensure that a) the clusters are spherical in shape with bounded diameter, while b) the time-complexity is less
than $O(n^2)$.

\begin{wrapfigure}[32]{R}{0.55\columnwidth}
  \centering
  \includegraphics[width=0.5\columnwidth]{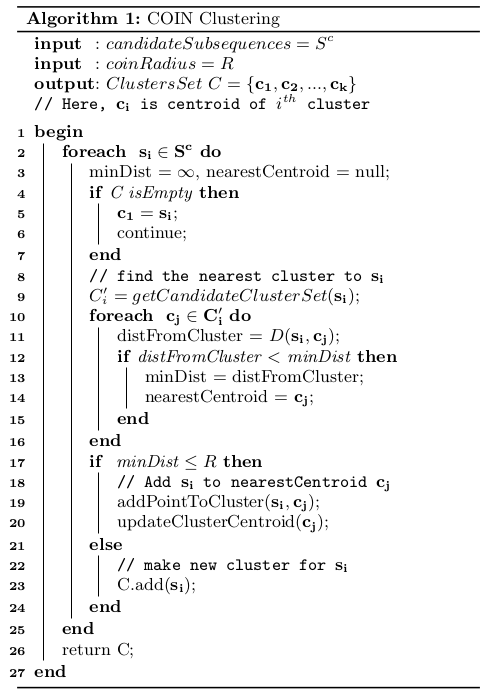}
  \label{a:basicCoinAlgo}
\end{wrapfigure}

According to Definition \ref{def:sim}, two time-series subsequences are
considered similar if the distance between them is less than $\delta$; so in COIN with threshold $R$, $\delta = 2R$. 
As a result, all subsequences 
within a cluster are guaranteed to be similar in that they are at most $2R$ apart according to the distance measure used.
\textbf{Note that $R$ can be taken to be a function of the variance of the un-normalized time-series and window length,
thus obviating the need to set $R$ as a parameter.}

\textbf{However, we shall assume that the entire series is normalized to have zero mean and variance one.}
Therefore, after normalization we shall take $R$ to be a function of the window length alone, as explained in Section 
\ref{sec:exp}.

Further, recall that we use the Euclidean distance measure between subsequences \textit{after} each is individually \textit{normalized} to
have zero mean. This normalization is required so as to detect multiple occurrences of the same waveform shape, albeit occurring at
different \textit{levels}. We shall recover multiple levels within a discovered frequent motif in Section \ref{s:motif}. 

For the moment we concern ourselves with clustering a set of subsequences $S^c$ obtained from the original series after series
normalization (\textit{and} also re-normalizing subsequences to zero mean before comparing them). 

Algorithm 1 shows the pseudo-code of a basic \textit{COIN Clustering} algorithm, which
is a variation of 1-NN (1-nearest neighbor) clustering where we compare every point with all the points clustered so far and find the nearest point. 
If the distance from the nearest point is less than a prior threshold bound, the new point joins the cluster of the
nearest point, otherwise a new cluster is created. 

The COIN Algorithm 1 takes coinRadius $R$, and $S^c$ the set of (normalized) subsequences to be clustered, as input.
(The difference between $S$ and $S^c$ will be made clear in Section~\ref{s:motif}.) It picks a point (i.e., subsequence) $\bf{s_i}$
from the set $S^c$, clusters it, and then picks the next point $\bf{s_{i+1}}$, and continues this until all of the points are clustered. To
cluster a point $\bf{s_i}$, it needs to find the target cluster for the point, efficiently. For this, it first determines a set of clusters that
are potentially near this point, such that the target cluster is almost sure to be in this set (line \# 9). We call this the
\textit{Candidate Cluster Set} $C_i'$; note that $C_i' \subseteq C$ (clusters found so far).

Next the algorithm compares $\bf{s_i}$ with the centroid of every cluster in Candidate cluster set $C_i'$,
and finds the nearest cluster-centroid $\bf{p_i}$ (line \# 10-16). If the distance between $\bf{p_i}$ and the point 
$\bf{s_i}$ is
less than $R$, the point $\bf{s_i}$ becomes a member of this cluster and the centroid is updated, otherwise a new cluster is created for the
point $\bf{s_i}$ (line \# 17-23).

From this description it seems that the properties (a) and (b) as mentioned above, are not completely achieved: 
The centroid of a cluster is updated every time a point joins the cluster; consequently some of the
points in the cluster may finally end up more than $R$ away from their centroids. 

In practice however, at least for multiple time-series in our vehicular sensor data, the number of such outliers was found to be less than $2k$, 
where $k$ is the number of clusters. We therefore chose not to re-cluster the outliers, and assert
that the clusters found by COIN are spherical in shape and have bounded diameter $2R$. 

Further, without re-clustering, the average time complexity of COIN algorithm is $O(k_a \times n \times d)$.
Here, $k_a$ is the average size of the candidate cluster set, $n = |S^c|$ is size of candidate subsequence matrix, and $d$ is the dimension
of each point (which in our case is determined by the subsequence window length $w$). Thus, as long as $k_a \ll n$ the complexity is only super-linear and an improvement over the quadratic behavior of both 1-NN
as well as earlier frequent motif discovery techniques (as we shall cover in Section \ref{sec:exp}).

In basic COIN clustering the candidate cluster set is taken to be all the clusters found so far, just as in 1-NN.
In practice, as shall also be evident from the results shown in Section \ref{sec:exp}, the number of clusters $k$ is very high. 
As a result basic COIN takes a long time to cluster even moderately large sets of time-series subsequences. 
To reduce the time-complexity we now describe schemes to reduce the number of candidate clusters.

\subsection{Coin Clustering: BIRCH acceleration}
BIRCH \cite{p:birch} is a well known bounded-spherical (COIN) clustering technique  that
stores clusters on leaf-nodes of a height-balanced tree, in which every node has at the most $B$ children. 
Every cluster is represented by its \textit{Clustering Feature}(CF) triple comprising of $(N, LS, SS)$. 
Here, $N$ is the number of points in that cluster, $LS$ is the linear sum of all the points in the cluster, and $SS$ is the squared sum of all the points in the cluster. 
The non-leaf nodes store the sum of CF triple of its children. Using the $LS$ and $N$ of CF triple, centroid can be calculated easily.
When looking up the target cluster of a point $\mathbf{s_i}$ (line\#9), in this tree, we start with the children of the root node and find
the nearest node $N_n$ among those, i.e. $\arg\min_j D(\mathbf{s_i}, e_j)$~$\forall$~children~$e_j$ of the root-node. 
Then we find the nearest child among the children of this node $N_n$, and go recursively down the tree to find the nearest leaf-node. 
This gives us the nearest cluster, and in this method of finding the candidate cluster set we return only one cluster (line\#9). 
It was observed empirically and is shown in Section \ref{sec:exp} that BIRCH gives significant gain in efficiency over
state-of-the art motif discovery techniques on our large vehicular data, thus making it usable in practice, 
while still being able to discover the same frequent motifs on standard data sets.
\subsection{Coin Clustering: LSH acceleration}

While COIN clustering of subsequences using BIRCH acceleration works well in practice, it appears to be an inherently serial technique. Further,
because of the nature of BIRCH that compares points with summary statistics of each node in its tree rather than the actual cluster
centroids,
it is often the case that a subsequence is \textit{not} placed in the cluster nearest to it. We now describe an alternative acceleration technique
that potentially addresses the above issues while remaining almost as efficient as BIRCH acceleration.

We use locality sensitive hashing (LSH) \cite{b:mmds} to create the candidate cluster sets for every subsequence. We
project subsequences random hyperplanes as suggested in \cite{p:IndykGaussianLSH}.
Each subsequence is hashed using $n$ random hyper-plane normals (the vectors $A$). As defined by the LSH technique \cite{b:mmds} 
we concatenate $r$ hash functions to get a bucket-id for each subsequence. We hash each subsequence and each cluster centroid
onto $b$ bucket-ids corresponding to $b$ different sets of $r$ hash functions.

Since the hyperplanes are random, it is highly likely that the target centroid $\mathbf{c}_i$ for a subsequence $\mathbf{s}_i$ falls in the same LSH bucket-id for
at least one of the $b$ sets of buckets. Conversely, we would like it to be highly \textit{unlikely} for centroids that are far away from $\mathbf{s}_i$ to share
a common bucket-id with this subsequence. We found that the values $r=3$ and $b=5$ work well in practice across datasets.

Once we have reasonable values of $r$ and $b$, the candidate cluster set is determined by hashing each cluster centroid as it gets created as
well as each incoming subsequence. Only those clusters are included in the candidate set whose centroids share at least one bucket-id with the incoming
subsequence. As we show in Section \ref{sec:exp}, COIN-LSH improves over COIN-BIRCH in clustering quality.

Further, since 
LSH partially pre-clusters the subsequences, albeit approximately, we could conceptualize a parallel implementation where subsequences hashed to
a single bucket-id are clustered in parallel, using paradigms such as map-reduce. Similar parallel algorithms for LSH-accelerated clustering
have been described in \cite{p:parallelER}, albeit in a very different scenario. We do not describe such a parallel version in detail; note
that because
there each subsequence is hashed to $b$ bucket-ids, so some post-processing will be needed to merge clusters, using techniques such as connected-components
as also described in \cite{p:parallelER}. However the analogy between these two scenarios is relatively easy to make, and serves as an
additional motivation
for COIN-LSH.

\section{Freq. Motif Discovery Process}
\label{s:motif}
\begin{wrapfigure}[27]{R}{0.7\columnwidth}
 \includegraphics[width=0.65\columnwidth]{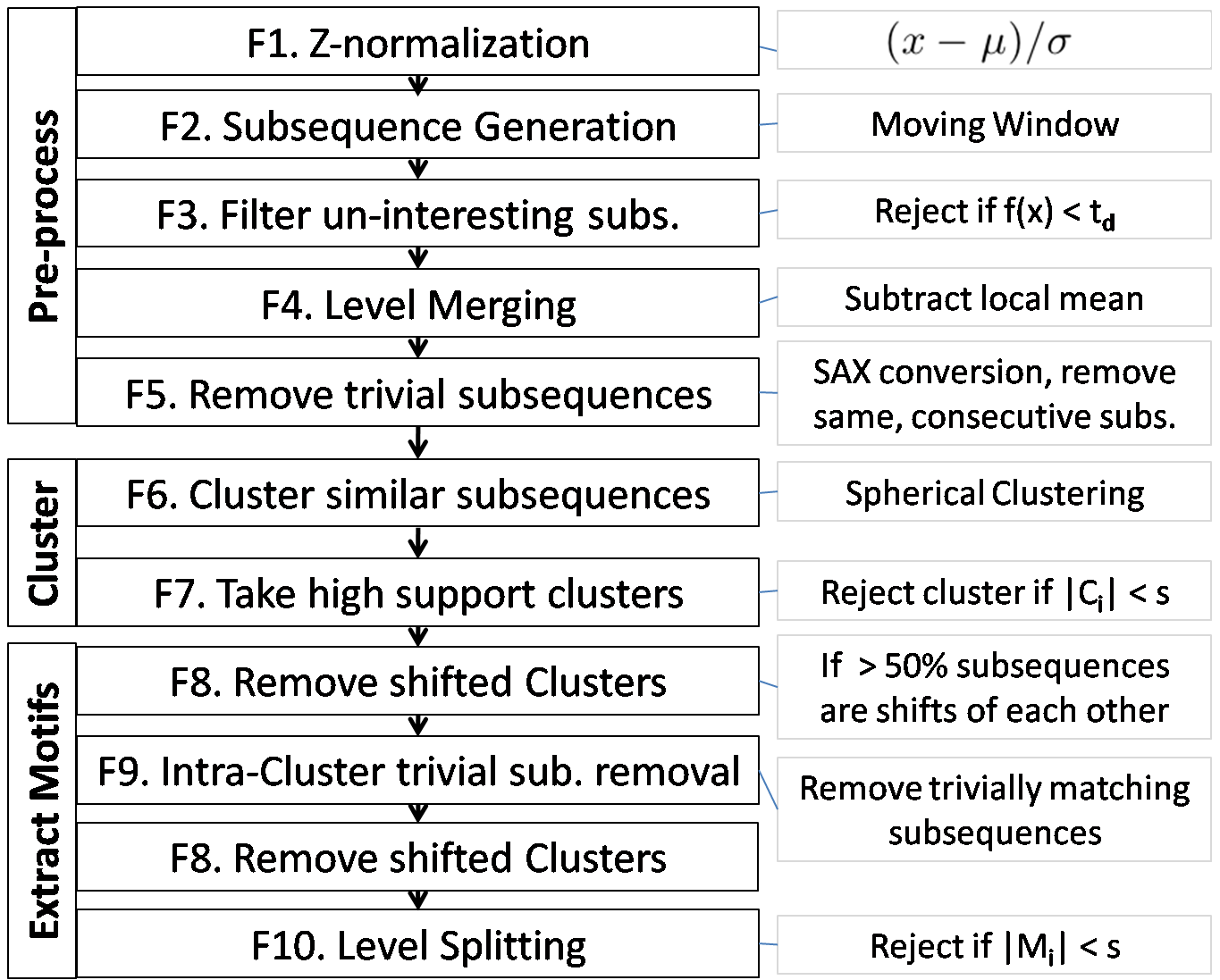}
 \caption{Complete frequent motif discovery process}
 \label{fig:overallMotif}
\end{wrapfigure}

\subsection{Pre-Processing \& Clustering (F1 - F7)}
We now describe the complete process of frequent motif discovery, see Figure \ref{fig:overallMotif}. 
(Some of these steps were also used in \cite{p:KeoghMotifs,p:KeoghExactMotifs,p:KeoghImageMotifs}.) 
Starting with the 
given set of time-series, we first \textit{normalize} (F1) them by calculating z-scores and \textit{generate
subsequences} (F2) using
a moving window of size $w$. 
We then \textit{filter uninteresting subsequences}(F3) where
subsequences in which the maximum deviation of z-score values is less than one ($\equiv$ the unnormalized series' standard deviation).

Next we \textit{merge levels} (F4) of these subsequences by first reducing the number of dimensions from $w$ to $d$, using piecewise
averaging. Then we subtract the local mean of each subsequence from it, so that all subsequences are of zero mean;
however, this step is reversed at a later stage of the process, i.e., F10.

In F5 we remove any consecutive subsequences that have the same symbolic representation according to the SAX symbolic representation \cite{p:KeoghSAX}. 
Neighboring subsequences having the \textit{same} symbolic representation are trivial matches: e.g., imagine a time-series segment of uniform slope; almost
all subsequences in such a segment are trivial matches and get removed by F5. The residual subsequence matrix, after \textit{removal of individually trivial subsequences} (F5), is called the \textit{candidate subsequence matrix} ${S}^c$.
Grouping subsequences with similar waveforms via COIN clustering, using either BIRCH or LSH acceleration is performed in F6;
however only \textit{high support clusters} ($|C_i| > s$) are retained (F7).

\subsection{Extract Motifs (F8 - F10)}
\label{sec:extract}
After taking the high-support clusters, in F8 we reject clusters that are shifted versions of other clusters, as per Definition \ref{def:shiftedClust}.
For this we proceed pair-wise, for each pair of clusters. Since the number of \textit{high-support} clusters is much 
smaller than the number of subsequences $n$ as well as the total number of clusters $k$, so this step, albeit of 
quadratic complexity, does not significantly affect performance.

First we sort subsequences in each cluster by their start-times.
We identify the cluster with smaller support ($H_1)$ and iterate over the subsequences of this cluster to match them with a few subsequences of the second cluster($H_2$).
We maintain two pointers that move on the ordered subsequences of the two clusters $\{H_1, H_2\}$ respectively.
In the beginning both pointers $\{q_1, q_2\}$ are placed at the first subsequence of respective clusters. 
We continue to move the pointer $q_2$ to next subsequence while 
$t(q_2) - t(q_1) < t_s$ (where the threshold $t_s$ is taken as $d$). In other words we are checking
if the two sequences occur within a window length of each other ($d$ is the reduced window length after piecewise averaging).

Of the subsequences over which $q_2$ is moved, we choose the one that is the nearest to the subsequence at pointer $q_1$ in cluster $H_1$. 
We then append the difference between the start times of this pair in a list hereafter referred as \textit{diff-list}. 
If however, there is no subsequence in $H_2$ that is less than $t_s$ away according to start-time for the subsequence starting at $q_1$ in $H_1$ we don't add any number to the \textit{diff-list}.
Next we move the pointer $q_1$ to next subsequence in $H_1$, and start $q_2$ with a few positions before current position (until
it finds a start-time $d$ less that of its current position). The procedure is repeated to find the nearest subsequence in $H_2$ to
the one at~$q_1$.

At the end, we count the number of items in \textit{diff-list} as well as its standard-deviation. Since this list contains
differences between
start times of nearest starting pairs between $H_1$ and $H_2$, the standard deviation is a measure of the correlation
between the start-times of nearby subsequences in each cluster. If the count is more than p\% of the size of $H1$ and the standard-deviation
of \textit{diff-list} is lower than $\sigma_t$, we conclude that the two clusters are near-duplicates, and we drop the smaller cluster $H1$. 
We have used $p=50$ and $\sigma_t = 2$ in practice.

%

\begin{wrapfigure}[9]{R}{0.5\columnwidth}
 \includegraphics[width=0.45\columnwidth]{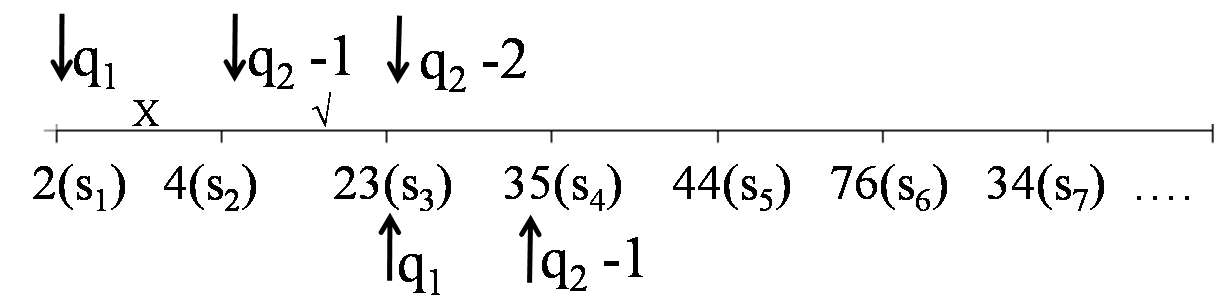}
 \caption{Intra-Cluster Trivial Subs. Removal}
 \label{fig:twoPointer}
 \end{wrapfigure}

Next we remove the trivially matching subsequences within each cluster (F9):
We again use a two-pointer algorithm, shown with an example in Figure \ref{fig:twoPointer}. 
Here, both of the pointers again move on the ordered subsequences but within a cluster only. 
Both pointers $\{q_1,q_2\}$ start with first subsequence and $q_2$ moves to next position. 
We consider all subsequences in $S$ (i.e., the \textit{original}, full set of subsequences) that start between $t(q_1)$ and $t(q_2)$, and if
we find subsequences $\mathbf{s_n}$ and $\mathbf{s_m}$ between $\mathbf{q_1}$ and $\mathbf{q_2}$
 for which $D(\mathbf{q_1}, \mathbf{s_n}) > \delta$ and $D(\mathbf{q_2}, \mathbf{s_n}) > \delta$ and $D(\mathbf{s_m}, \mathbf{s_c}) > \delta$, (here $\mathbf{s_c}$ is the centroid of the cluster)
we consider the subsequences at $q_1$ and $q_2$ as non-trivially matching, see Definition \ref{def:sim}.
If however no such subsequence is found between the two pointers, we reject subsequence at $q_2$ and move further.

Finally, since some of the subsequences have been removed the support of each cluster may change, and it becomes possible
that some shifted clusters that failed detection during F8 can be discovered by re-running F8. The clusters of subsequences obtained after
this stage are referred as \textbf{group-motifs}.

 \begin{wrapfigure}[15]{R}{0.77\columnwidth}
  \centering
 \includegraphics[width=0.75\columnwidth]{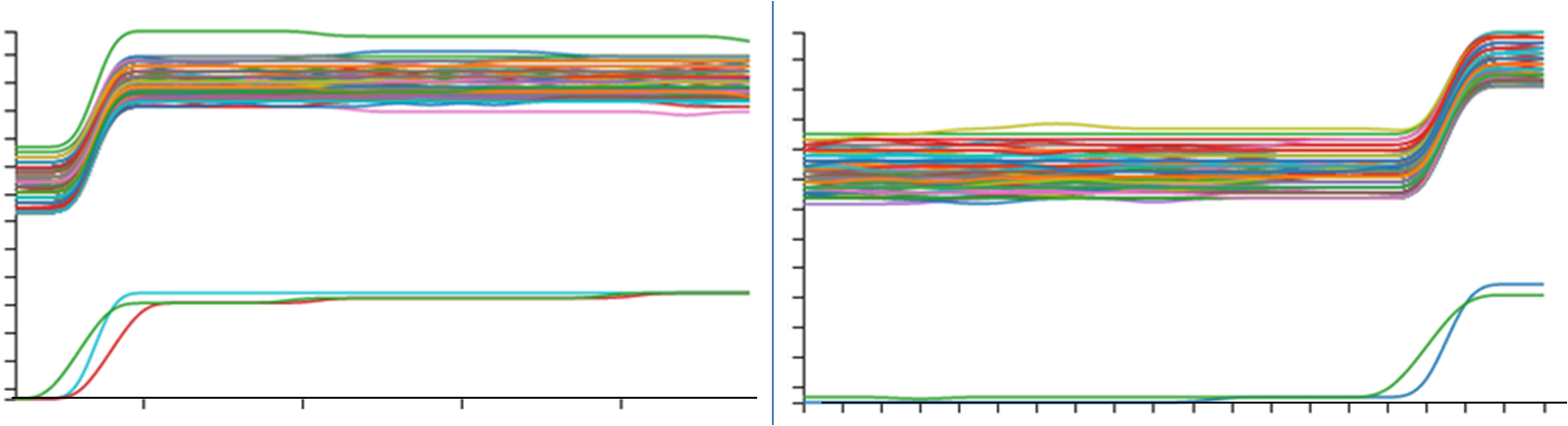}
 \caption{Charts show time on x-axis and sensor value on y-axis, and demonstrate the motivation for level splitting.}
 \label{fig:split}
 \end{wrapfigure}
Figure \ref{fig:split} shows a sample of the group-motifs obtained after this stage 
it can be observed that these patterns happen at different levels of magnitude, and should not be clubbed
in the same motif for practical purposes: For example, it does not makes sense to consider a `sudden rise' in temperature 
from zero to 50 degrees in the same light as a rise, albeit equally sudden, from 100 to 150 degrees; the former may be quite natural, 
while the latter might result in system failure.
Therefore, \textit{level splitting} (F10) of these group-motifs is performed; this undoes the effect of step F4 (mean-shifting within subsequences)
that resulted in shapes occurring at different levels being clustered together.

For level-splitting we first calculate \textit{level} of every subsequence in a group-motif as its \textit{mean} $L(\mathbf{s_i}) = \frac{1}{w} \times
\sum_{j=0}^{w} \bar{s}^j_i$. (Here we use the original
subsequence $\mathbf{\bar{s}_i}$ \textit{before} zero-mean normalization.) To split a group-motif according to levels, we run
one-dimensional DBSCAN clustering on the levels of subsequences in a motif, and split the group-motifs according to the clusters found.
Again, only the motifs that have support greater than $s$ are chosen as final motifs. 

Finally, we are left with the final frequent motifs $\mathcal{M} = \{M_1,M_2,...\}$ from the set of time-series $T$, segregated by level. 
In summary, our end-to-end frequent-motif discovery algorithm takes only \textit{three} input parameters, the radius ($R$), support $s$ as used in
Definition \ref{def:motif}, and window length $w$.

\section{Experiments and Analysis}
\label{sec:exp}
\subsection{Datasets and Infrastructure}
We report the experimental results on vehicular sensor data for 59 different runs of various vehicles. All these vehicles
were driven for almost 3 hrs. When these vehicles were being driven, values of 27 different sensors were recorded continuously, while we
report results of experiments on select 6 sensors only. This lead us to our primary dataset for this paper, which contains $m=59$
time-series for every sensor. These sensor readings were taken at a regular interval of 1 second; consequently, number of subsequences for a
window length of $w=20$ were almost $650k$. In addition to the vehicular sensor data, we also performed experiments on some publicly
available datasets in order to facilitate the comparison of our approach with those available in research literature. These public datasets
are Electrocardiogram, Bird-calls, and Temperature data \cite{d:epenthesis}. 

For our COIN-BIRCH algorithm, BIRCH code for step F6 was taken from \cite{o:birch}.
We have made our end-to-end code available at \cite{o:coinMotif} for
others to easily verify our results.
We also publish subset of our vehicular sensor data at \cite{o:coinMotif}.
Code for the Epenthesis algorithm described in \cite{p:epenthesisMotif} was taken from \cite{d:epenthesis}, its approach is summarized in
Section \ref{sec:rw}.

All efficiency related experiments were performed on a machine with processor: Intel Xeon E7520@1.87GHz, 4
physical CPU of 4 cores, configured with 32-virtual processors, and RAM: 32GB, while no other process was running on this computer.

\begin{figure*}
  \centering
  \includegraphics[scale=0.6]{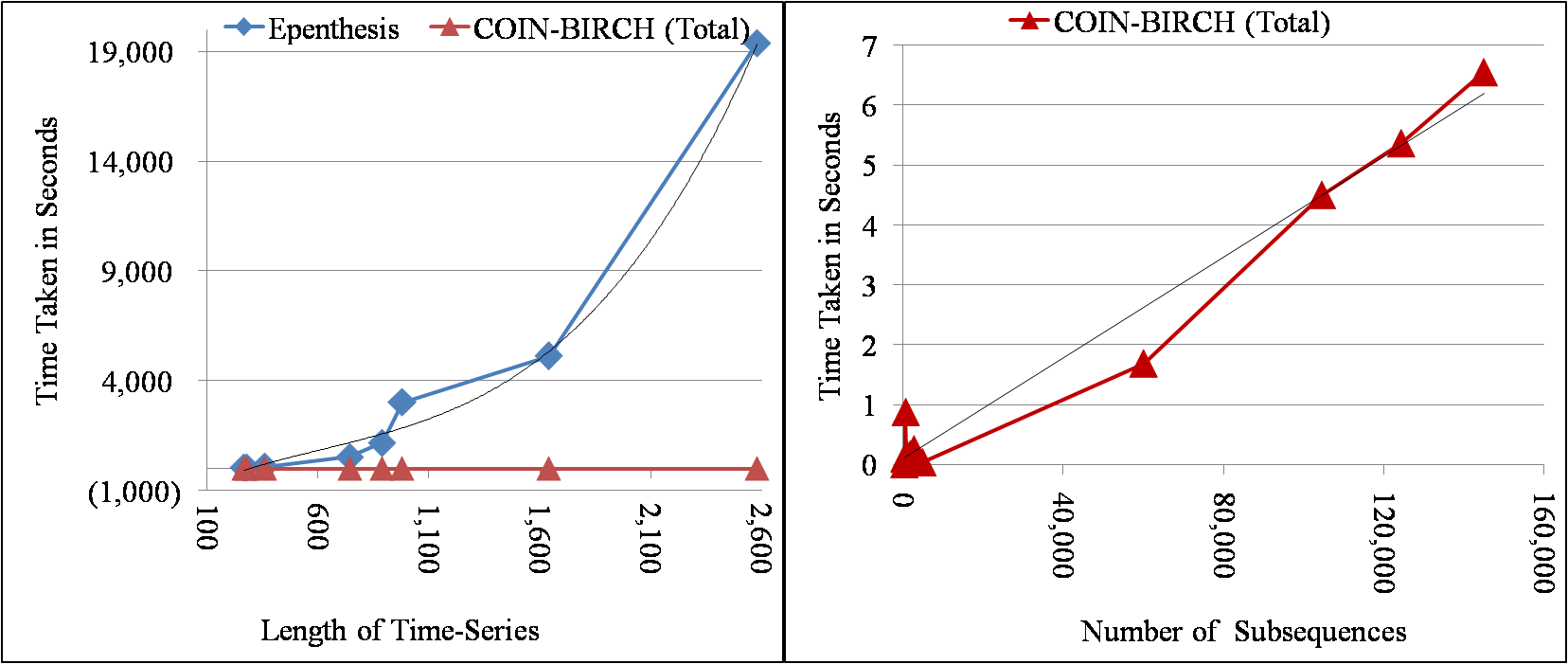}
  \caption{Efficiency comparison of COIN-BIRCH and Epenthesis for fixed window length}
  \label{fig:eff}
\end{figure*}

\subsection{Parameter values for experiments}
Our method of discovering frequently occurring motifs, takes $w, R, s,$ and $f$ as input parameters, however except for $w$, suitable values
of all the parameters can be derived from data statistics. In this section we describe how the values of these parameters was chosen for
various experiments. 

Before clustering the subsequences, we z-normalize the whole time-series. The first parameter $R$ binds the size of the clusters, i.e., it
is a measure of how far apart can various subsequences be within a motif. In our experiments we found that mostly $R=1$ works for 
$w=20$ in all time-series (that we experimented with), and used the same value unless otherwise stated. Further, when discovering the
motifs with width $w > 20$, we adjust R as $R_{adj}=R \times \sqrt{w/20}$. This strategy was observed to work well for motif widths
upto $w=200$, and did not test for motifs wider than $200$.

Further, in vehicular time-series data from multiple runs of vehicles, we used minimum support $s = 50$ (see F7 in Section \ref{s:motif})
since there are 59 runs 
and we want patterns that occur in almost all runs. For public datasets we used minimum support $s=2$ since all of these datasets are small
and can be observed through simple time-series visualization techniques see Figure~\ref{fig:pub1}. We filter all subsequences, (see F3 in
Section \ref{s:motif}) which have normalized deviation (difference between max and min values, in a normalized subsequence) smaller than
$f=1$.

\subsection{Results}
In vehicle sensor data we used $w=20$ based on inputs from domain experts. We also found motifs of wider lengths, however discovering
suitable time-series window-length in linear time still remains a research challenge. For $w=20$, the size of our subsequence
matrix was $\approx 650k \times 20$. After piecewise averaging, the dimension was reduced to $d=10$, i.e., the matrix size $\approx 650k
\times 10$. The size of the candidate subsequence matrix $S^c$ obtained after pre-processing(F1-F5) varied from almost $600$ to $170k$,
for different sensors, since some subsequences are removed either because they show little deviation or represent trivial-matches. 

The number of clusters found varied from almost $350$ to $170k$. 
However, only $0$ to $400$ of these clusters pass the minimum-support condition. 
Further, on an average $\sim$25\% of the clusters were dropped when we removed shifted versions of the clusters, and we obtained about 1 to
250 clusters.  Next, during the removal of trivial subsequences, $\sim$17\% of the subsequences got dropped.

\begin{figure*}
\centering
 \includegraphics[width=185 mm]{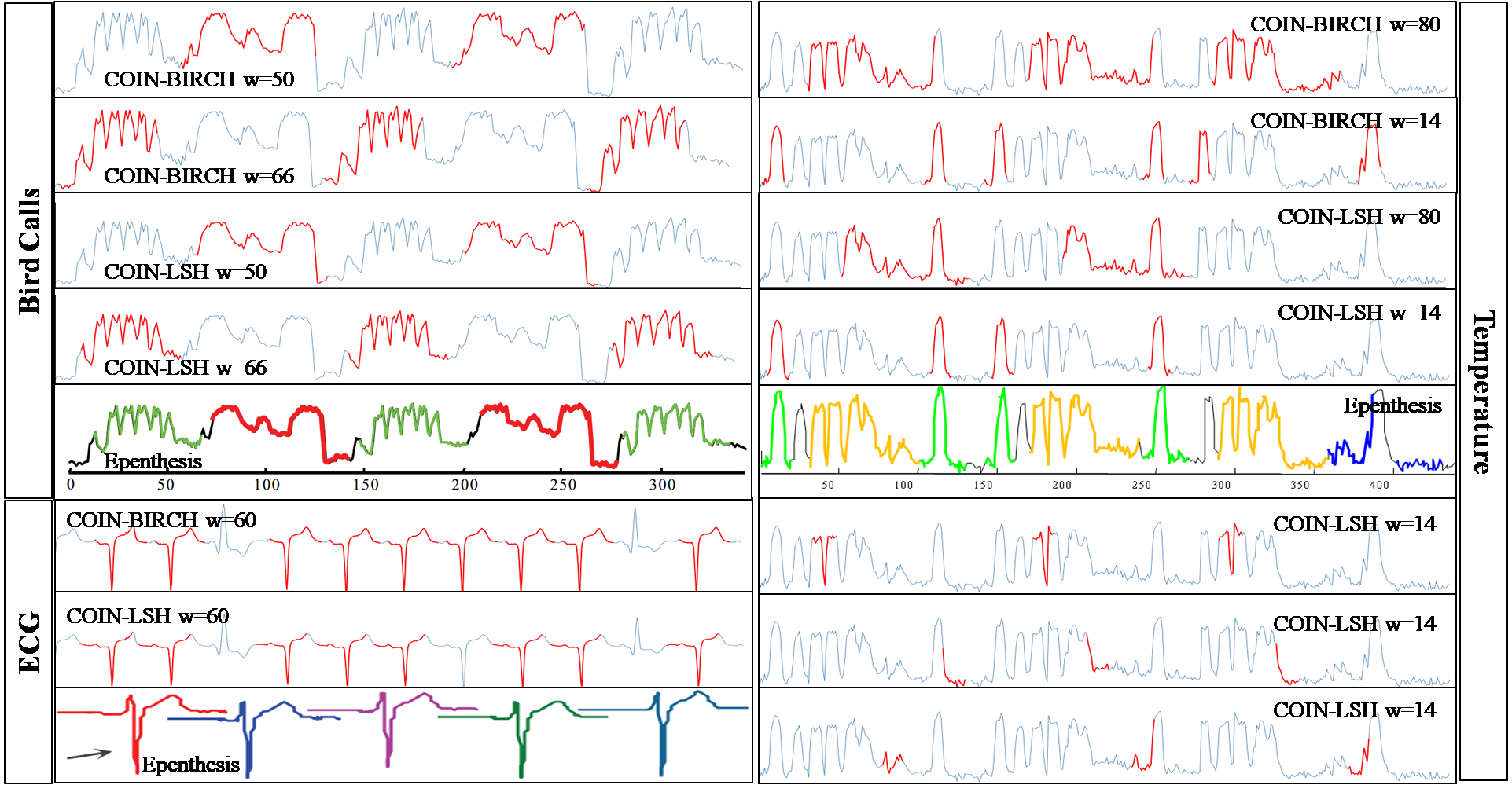}
 \caption{Motifs on public datasets, pictures marked by ``Epenthesis'' taken from \cite{p:epenthesisMotif,d:epenthesis}}
 \label{fig:pub1}
\end{figure*}

\begin{wrapfigure}[13]{R}{0.55\columnwidth}
  \centering
  \includegraphics[width=0.5\columnwidth]{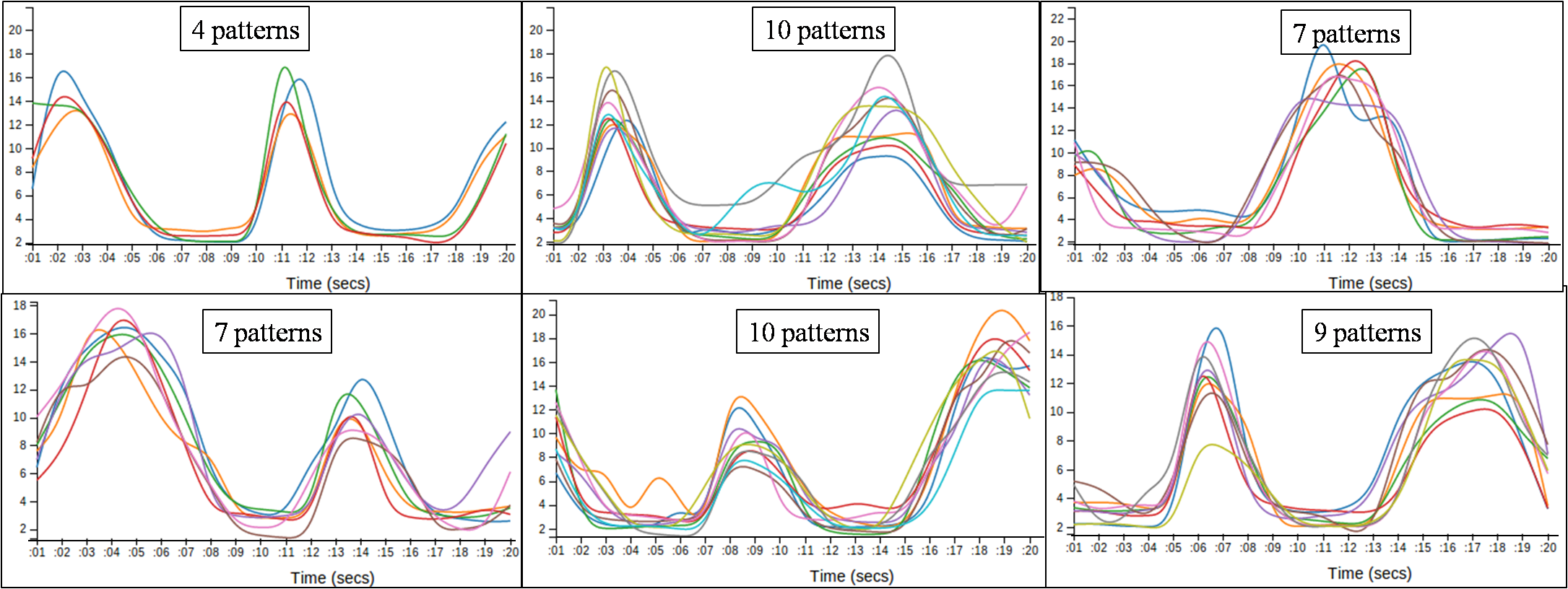}
  \caption{Motifs Detected on Sample Vehicle Sensor Data}
  \label{fig:vsMotif}
\end{wrapfigure}

Figure \ref{fig:eff} shows the running time of COIN-BIRCH as well as the Epenthesis algorithm
for time-series of different lengths. As evident from the plot on the left, COIN-BIRCH is significantly more efficient
than Epenthesis, which iteratively finds pair-motifs to create clusters instead of subsequence clustering. 
Further, the plot on the right of Figure \ref{fig:eff} shows that COIN-BIRCH works efficiently even as the number of subsequences
grows well beyond the capacity of techniques such as Epenthesis, as was the case for our vehicular sensor data. Motifs 
discovered from our sample vehicle sensor data, which has been shared at \cite{o:coinMotif}, are shown in Figure 
\ref{fig:vsMotif}.

We also tested our algorithm on public datasets, e.g., Electrocardiogram(ECG), BirdCalls, and Temperature datasets
\cite{d:epenthesis}. These experiments took less than 2 sec to run, and some of the motifs discovered on these
datasets are shown in Figure \ref{fig:pub1}, demonstrating that our techniques discover the same motifs as Epenthesis
\cite{p:epenthesisMotif} on public datasets. On BirdCalls data our
algorithm also discovers the similar motifs as done by Epenthesis. Note that while COIN-LSH did discover the heart-beats in the ECG data, it
missed some of the instances of the motifs and the COIN-BIRCH did find all of the instances of the motif. Further, on Temperature data for
motif width $80$, we used a COIN radius $R=2$. COIN-BIRCH in this case also found all the motifs as discovered by Epenthesis, while COIN-LSH
found $6$ more motifs for window length of $w=14$. All of these actually are the real motifs according to our definition, 3 of these are
shown below the Epenthesis image on the right hand side in Figure \ref{fig:pub1}. Results on other public datasets are also very similar
and can be verified through our code available at \cite{o:coinMotif}.

In COIN-LSH we used $r=3$ and $b=5$. 
We found both COIN-BIRCH and COIN-LSH give similar
performance, although COIN-BIRCH was usually faster (of course COIN-LSH
is more amenable to parallel implementation, as mentioned earlier). 
The above experiments were also tried for different window lengths and random order of clustering of subsequences in $S^c$, with similar
results.
\section{Related Work}
\label{sec:rw}
Motif discovery from time-series data has been an active topic of research in last decade or
so \cite{p:KeoghMotifs,p:moen,p:KeoghExactMotifs,p:epenthesisMotif,p:effiMotif}. It is also evident from past work that
there are many aspects of this problem which need to be addressed, for example \cite{p:epenthesisMotif,p:moen,p:effiMotif} focus on finding
out what should be the appropriate width of the time-series motifs, and find the motifs of multiple lengths. Most of these
approaches use MK-Motif discovery algorithm \cite{p:KeoghExactMotifs} underneath, which discovers pairs of subsequences that are similar.
Xi et al. in \cite{p:KeoghImageMotifs} focus on finding another subsequence that is similar to a given subsequence.

Our definition of frequent time-series motifs is similar to that of Keogh et al.
\cite{p:KeoghMotifs,p:epenthesisMotif}; however they do not focus on efficiency. 
These approaches have quadratic \cite{p:KeoghMotifs} or cubic \cite{p:epenthesisMotif} time-complexity
in the size of the series ($|T|$) (recently, \cite{p:effiMotif}\footnote{Source code not published.} brings \cite{p:epenthesisMotif} down to
quadratic complexity). 
Chiu et al. \cite{p:KeoghMotifs} exploit symbolic representation of subsequences using the SAX scheme 
\cite{p:KeoghSAX}, 
we directly use subsequences in $R^d$ space after z-normalization and level-merging. 
In \cite{p:epenthesisMotif} authors exploit pair motif discovery algorithm \cite{p:KeoghExactMotifs} followed by a search for other similar
subsequences in the time-series. They choose the member subsequences of a frequent motif based on bits saved through 
MDL encoding of the subsequences of the motif. We improve on these approaches, empirically achieving near linear 
performance; see Section \ref{s:coin} and \ref{sec:exp}. 

Subsequence clustering (STS) was identified as a research challenge by Keogh et al. in \cite{p:keoghClustMeaning}. They demonstrated that
a)~the output of STS clustering is independent of the dataset used to generate them and that b)~subsequences contained in a cluster
don't share the same waveform and therefore lead to smoothing effect, resulting in sinusoidal motifs being detected for all
time-series. This was demonstrated through the use of k-means and hierarchical agglomerative clustering. However, Goldin et al.
\cite{p:meaningful}
demonstrated (through the use of another distance measure) that the output of STS clustering has correlations with the datasets used.
Denton et al. showed (through the use of kernel-density based clustering) that in 7 out of 10 cases 
there is a correlation between the clusters and the datasets used. 
Minnen et al. in \cite{p:meanShiftMotif} also use density-based clustering of non-overlapping subsequences. 

Detailed analysis of the challenges involved in STS clustering was presented by Chen \cite{p:ChenICDM,p:ChenKAIS}. 
He proposed an alternate distance measure to solve this issue. We submit that the use of bounded spherical, i.e., COIN clustering for
discovery of frequent motif from time-series works in practice, so STS clustering is meaningful, at least to us as, we found it
useful as well as highly efficient for our practical application scenario. Code for these techniques was not
available; however, they did not focus on efficiency per se, and used standard clustering techniques such as k-means
that is clearly outperformed by BIRCH as shown in \cite{p:clusteringSurvey1}.
Of course, unlike the Epenthesis approach of \cite{p:epenthesisMotif} that is parameter free, approaches
based on subsequence clustering all rely on at least the motif width being an input parameter.

To the best of our knowledge \cite{p:clusteringSurvey1},
little has been written regarding the use of bounded spherical (COIN) clustering techniques, especially 
for motif discovery, e.g., using BIRCH \cite{p:birch}. Our COIN-LSH
approach improves on quality of motifs discovered using BIRCH while showing similar performance,
and is also parallelizable using techniques such as in  \cite{p:parallelER}.
A concept similar to LSH was used in \cite{p:KeoghImageMotifs} for discovery of pair motifs in images, 
but on discrete symbolic representation of time-series, and the hash-functions were chosen by omitting specific dimensions. 
In contrast, we use subsequences in their original form and hashing based on random hyperplanes in d-dimensional space. 
A concept similar to COIN has been used in \cite{p:KeoghExactMotifs}, but for pair motifs rather than frequent motifs.

The problem of trivially matching subsequences has been identified in the research literature related to STS clustering
\cite{p:ChenICDM,p:ChenKAIS,p:useful,p:KeoghMotifs,p:keoghClustMeaning,p:meanShiftMotif}. Most of these approaches
\cite{p:ChenICDM,p:ChenKAIS,p:meanShiftMotif} focus on non-overlapping subsequences at the outset therefore such
approaches may altogether miss some of the motifs due to their lower support. 
Further, Chen has also argued in \cite{p:useful} that removing the subsequences before clustering also does not completely solve the issue
of smoothing of subsequence clusters. Not enough attention has been given to an approach for removal of such subsequences after clustering, 
primarily because of the absence of suitable clustering method itself. 
Similar to above publications we also remove the trivially matching subsequences before clustering, however one of the
key-contributions of our work is removal of trivially matching subsequences through post-processing, see Section~\ref{s:motif},
as well as highlighting the importance of level-splitting so that the discovered motifs are useful in practice.

Finally, note that our definition of frequent motifs is very different from those presented in
\cite{p:KeoghImageMotifs,p:KeoghExactMotifs,p:ClustTSShapelets} as they either focus on finding pairs of similar
subsequences or clustering different time-series rather than subsequences of the same series (so they do not face
the problem of trivial matches).

\section{Using Motifs Further}
\label{sec:applications}
As we mentioned at the outset, motifs are merely one potentially useful element in the overall process of sensor data
analysis.
For example, one of the goals of our analysis is to predict the occurrence of faults in a vehicle as early as possible. If one
considers a sensor's historical behavior preceding a fault, within a reasonable time frame (e.g. hours or at most days, typically with
a gap excluding the segment immediately preceding the fault), as a `positive' example, and other similar length histories not close to a fault 
as `negative' ones, we can view the fault prediction problem as one of time-series classification. 

Frequent motifs, especially those that are also frequent \textit{within} the set of positive histories, are \textit{one} of the features
used while training a classifier for early detection of faults, or for rule-based techniques such as \cite{vilalta2002predicting}. 
Frequent motifs, as also discretized levels of the sensor value, its derivative, etc. as well as other features 
can also be combined for such classification tasks: Keogh \cite{xing2010brief} provides a survey of time-series, and more generally 
sequence classification techniques.
\section{Conclusion and Discussion}
\label{sec:concl}
We first highlight key aspects of our algorithm and then present a summary of what has been presented in this paper.
Firstly, even with our best effort we could not find any other publication that addresses the same problem as ours directly, except
\cite{p:KeoghMotifs}, which is clearly outperformed by our approach from the efficiency perspective.
For further details see Section \ref{sec:rw}. For the same reason, almost all
publicly available datasets are not appropriate to demonstrate the usefulness of our approach since they are all far too small. 
We have made available the end-to-end source code of our approach\cite{o:coinMotif} along with
a comparatively bigger time-series extracted from our vehicular sensor data (albeit still far smaller than our real-life data).

Further, it has been advised in \cite{p:keoghClustMeaning} that in order to solve the problems of trivial matches and smoothing effect, some
of the subsequences should be removed before clustering; we also remove selected subsequences before
clustering (see Section \ref{s:motif}). It can be observed from Figure \ref{fig:pub1}, i.e., the results on ECG data that our approach
sometimes
misses a few instances of the motifs. However we do detect the frequently occurring waveform patterns, which is the primary goal of motif discovery.
Once a pattern has been discovered, finding all instances of a particular pattern, including any missed during the clustering process,
can be found by subsequence matching in a linear scan of the time series.

A potential criticism of our approach can be that it requires setting an input parameter, namely the window length.
However, the parameter-free approach such as \cite{p:epenthesisMotif} is too computationally expensive to run on our large real-life
vehicular sensor data.
Note that remaining parameters such as the radius are derived from this single input parameter and/or the statistics of the time-series.

Finally, it may be argued that comparing our approach with \cite{p:epenthesisMotif} is unfair in that the goal of the latter is not efficiency,
and there are other published works on motif discovery that also use subsequence clustering. Nevertheless, \cite{p:epenthesisMotif}
was the only publicly available implementation for discovering frequent motifs in time series. Further, previously published approaches using
subsequence clustering were also of quadratic complexity and therefore did not appear to be worth re-implementing for our purposes as they
were unlikely to work on large data. 

We have shown that useful frequent time-series motifs can be discovered using COIN clustering of time-series subsequences, in which
spherical clusters are tightly bounded by a threshold. We have also addressed additional important practical aspects of 
the frequent-motif discovery problem, of which clustering is only a part. 
These include the removal of trivially matching subsequences as well as the elimination of near-duplicate motifs that are merely 
shifted versions of each other. We have presented experimental results on real-life vehicular sensor data as well as
on public data sets. 

Our algorithms are efficient, i.e., near linear time in the size of the series, making them
useful in practice with big data. We have shown that our technique discovers meaningful frequent motifs, is significantly more efficient
than state-of-the art techniques such as \cite{p:epenthesisMotif} and was found usable in practice on large collection 
of vehicular sensor data. We have published our code along with an extract from our real-life dataset for others to use 
or reproduce our results.

\bibliographystyle{abbrv}
\bibliography{99-PuneetMisc,95-PuneetPapers,90-iLabDelhi,70-TS-Motifs,75-Clustering,50-MapRedHad,98-Data,seqclass}

\end{document}